# Hedonic Models of Real Estate Prices: GAM and Environmental Factors

Jason R. Bailey [1], Davide Lauria [1], W. Brent Lindquist [1], Stefan Mittnik [2], Svetlozar T. Rachev [1]

[1] Department of Mathematics & Statistics, Texas Tech University, Lubbock, TX 79409-1042, USA;
   jason.r.bailey@ttu.edu; davide.lauria@ttu.edu; zari.rachev@ttu.edu

[2] Scalable Capital, 8053 München, DE; stefan@scalable.capital

**Abstract.** We consider the use of P-spline generalized additive hedonic models for real estate prices in large U.S. cities, contrasting their predictive efficiency against linear and polynomial based generalized linear models. Using intrinsic and extrinsic factors available from Redfin, we show that GAM models are capable of describing 84% to 92% of the variance in the expected ln(sales price), based upon 2021 data. As climate change is becoming increasingly important, we utilized the GAM model to examine the significance of environmental factors in two urban centers on the northwest coast. The results indicate city dependent differences in the significance of environmental factors. We find that inclusion of the environmental factors increases the adjusted $R^2$ of the GAM model by less than 1%.

## 1. Introduction

Real estate prices are often analyzed using hedonic models to capture the heterogeneous effects of factors that are both intrinsic (to the residence) and extrinsic (in a broad sense, i.e., to the location). Hedonic models use regression to quantify the relationship and effect of each factor on the price of a residence. The development of a model generally comprises three steps: (1) identification of relevant factors; (2) selection of a regression formulation; and (3) application of the model to real-world data.

Generally accepted intrinsic factors include: the number of bedrooms and bathrooms; indoor and outdoor areas (square footage); and the categorization of the dwelling type (single-family, condominium, etc.). Belke and Keil (2017) established the validity of several macroeconomic factors through a panel study of German regions. Extrinsic macroeconomic factors that were identified included but were not limited to: the number of newly constructed apartments per one thousand inhabitants of each city; the recorded number of real estate market transactions per one thousand inhabitants of each city; the unemployment rate in said cities; the purchasing power index of the area; and the number of hospitals - used as a proxy for the city's overall quality of infrastructure.

Significant progress has been made in examining further extrinsic location-related measures as explanatory factors for real estate prices. Postal codes are often correlated with factors related to neighborhood desirability, and GPS coordinates can provide precise location measurements with more granularity. Hill and Scholtz (2018) demonstrated the superiority of a nonparametric spline surface based on GPS data over postal code proxy information as a way of controlling for locational effects. Indeed, many publicly available geocoding websites can provide the latitudinal and longitudinal coordinates of an estate with speed and accuracy; such refinements have allowed for more expansive analyses. Helbich et al. (2013) studied the explanatory power of exposure to solar radiation on the pricing of owner-occupied flats in Vienna by employing airborne LIDAR maps. Olszewski et al. (2017) studied the effects of time, housing policy, and spatial relationships on housing prices and verified the significance of such factors as the distances to the nearest metro station, green spaces, and the city center. Cohen and Coughlin (2008) studied the effects of home proximity to airports. Their work confirmed that homes close to Hartsfield-Jackson International Airport in Atlanta that experienced significant noise levels had lower selling prices than equivalent homes without the noise levels. Interestingly, homes which were close to the airport but without the high noise levels had higher selling prices than equivalent homes further from the airport, suggesting that appropriately located proximity to an airport is an amenity. Colonnello et al. (2021) considered a linear hedonic model for housing yield (rent-to-price ratio), incorporating a relatively large number of extrinsic, demographic and local economic factors.

As hedonic models aim to estimate the contributory value of each internal or external factor, the decomposition allows for the appropriate use of generalized linear, additive, or logarithmic models to identify the contributive power of each factor. Pace (1998) was one of the earliest to employ a generalized additive model (GAM) in the context of real estate pricing, demonstrating that GAMs could outperform more simplistic parametric and polynomial models. Owusu-Ansah (2011) presented a review of parametric, non-parametric, and semi-parametric models and summarized the strengths and weaknesses of each approach. Silver (2016) proposed a hedonic regression pricing methodology that combined "time dummy," "characteristics," and "imputation" hedonic approaches. He argued that the methodology mitigates substitution bias, accommodates thin markets, requires only periodic regressions for reference periods, and is not subject to data misspecification and estimation issues. Using a structured additive regression (STAR) model, Brunauer et al. (2013) regressed individual attributes and locational characteristics through a four-level hierarchical model to quantify the contribution of each level of geographic detail to housing prices. For example, the level-2 categorization (municipality) captured macroeconomic housing policies, while the level-4 categorization (county) captured county economic policies (e.g., property taxes) even if the policies at either level were not explicitly identified. Bárcena et al. (2013) employed a semi-parametric and geographically weighted hedonic model to create an index of housing prices in Bilbao, Spain over the time-period before and after the Great Recession. In doing so, they were able to identify the impact of commonly accepted factors (e.g., garage presence, city district) on housing prices and produce a model whose results produced improved agreement with a price index produced by a governmental institute. Bax and Chasomeris (2019) employed a generalized linear model (GLM) to measure apartment rent prices from a set of statistically significant factors, which included the floor area, number of bathrooms, number of bedrooms, and the name of the suburb to which the apartment belonged. Eiling et al. (2019) used monthly housing returns for 9,831 zip codes across 178 U.S. Metropolitan Statistical Areas (MSAs) to quantify the systematic market risk and idiosyncratic zip-code specific risk within each MSA. Their findings show that systematic risk and idiosyncratic risk were both positively priced in over 20% of all MSAs and that both results were related to liquidity levels in the housing markets.

The acronym ESG (environmental, social, governance) refers to the sustainability factors of a property. Examples of environmental factors include but are not limited to: usage of renewable energy, the reuse of water, the residence's ability to withstand and adapt to increased temperatures as a consequence of global warming, and the risks of natural disasters. Examples of social factors include, but are not limited to: customer satisfaction, employee (i.e. construction worker) satisfaction, labor standards, and noise issues. Examples of governance factors include but are not limited to: transparency in the company and/or owner, presence of legal issues or corruption in the company and/or owner, and compliance with regulations at the local, state, and federal levels. As we progress further into the 21st century, ESG factors will in undoubtedly take higher priority in the valuation and construction of residences.

In this paper, the effectiveness of a GAM pricing model is contrasted against several GLM models for five U.S. cities. For three of these cities, New York City (#1 in U.S. population in 2021), Los Angeles (#2) and Louisville, KY (#56), we consider seven standard intrinsic and four (five for New York City) extrinsic price factors. For two additional cities, Seattle (#18) and Portland, OR (#25), four additional factors representing environmental considerations are considered. For these two cities, we investigate to what extent environmental factors are significant in real estate pricing, and to what extent their inclusion improves overall accuracy.

## 2. Materials and Methods

*2.1 Response and Factor Data*

Except as noted at the end of this paragraph, price and factor data were obtained from Redfin, a leader in online real estate listings.[1] Our data set comprises housing offers in New York City (NYC); Los Angeles (LA); Louisville, KY; Portland, OR; and Seattle as listed at the end of the second quarter of 2021. Our data set comprised dwelling price and ten factors: dwelling type (single-family homes, multi-family homes, townhouses, and condominiums); the number of bedrooms (Beds); the number of bathrooms (Baths); living area (Indoors); lot size (Lot); the year during which the construction of the dwelling was completed (Year); the number of days on the market (Days); the monthly homeowners' association fee (HOA); latitudinal (Latitude); and longitudinal (Longitude). In the case of NYC, the data also includes the borough to which the property belongs. Additionally, we collected the locations of local sex offenders (familywatchdog.us), which we converted into latitudinal and longitudinal coordinates from which the distance to the nearest listed residence was computed (Distance).

---

[1] Data from redfin.com was collected by specifying the city and the entries for "All filters". The specific filter values are provided in the appendix.

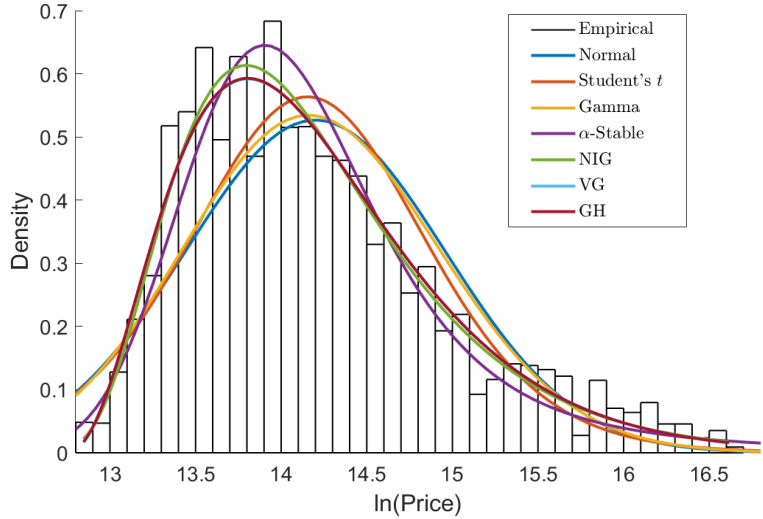

**Figure 1.** The fits (solid curves) of alternative distributions to the empirical log-price data (histogram) for NYC. (GH: generalized hyperbolic, VG: variance gamma, and NIG: normal-inverse Gaussian).

Our data set includes three factors that are often not considered: days on the market, the presence of HOA fees, and distance to the nearest sex offender. The latter is largely a factor of U.S. focus; many countries do not make the location of sex offenders publicly available. HOA fees were included to capture microeconomic factors. We included days on the market to capture aspects of consumer subjectivity relative to real or perceived reasons why a home might be on the market for a significant number of days beyond the average.

Due to the heavy-tailed nature of dwelling prices, we used ln(Price) to express dwelling price (log-price). Fig. 1 displays the empirical distribution of the log-prices for the NYC data. Also shown are best fits to the log-price empirical distribution using symmetric and skewed distributions. The generalized hyperbolic (GH) and normal-inverse Gaussian (NIG) distributions provided the best fits to the log-price data.

*2.2 Generalized Models: Additive and Linear*

We employed the GAM model (Hastie and Tibshirani, 1990),[2]

$$g(\mu_i) = \beta_0 + f_1(x_{1i}) + f_2(x_{2i}) + \cdots + f_m(x_{mi}), \qquad (1)$$

for the expected value $\mu_i = E[Y_i]$ of the univariate log-price response variable $Y_i$ in terms of the intrinsic and extrinsic factors $x_i$, $i = 1, \ldots, m$. It is assumed that $Y_i \sim EF(\mu_i, \theta)$, where $EF(\mu_i, \theta)$ denotes the exponential family distribution with mean $\mu_i$ and scale parameter $\theta$. The function $g(\cdot)$ is referred to as the link function, as it relates conditional expectations of the log-price to the factors via

$$\mu_i = g^{-1}\big(\beta_0 + f_1(x_{1i}) + f_2(x_{2i}) + \cdots + f_m(x_{mi})\big).$$

---

[2] Specifically, we utilized the *gam* function from the R package *mgcv*. P-spline basis functions and the identity link function were specified. Aside from specifying the functional form, all other *gam* arguments were set to default values.

As conditional expectation for the best-fit GH and NIG models is in the domain of attraction of the normal distribution, we used the identify function for $g(\cdot)$. For the functions $f_j(\cdot)$, we used P-splines (Eilers and Marx, 1996), which minimize the penalized sum of squares

$$\sum_{i=1}^{N}\left(Y_i - \sum_{j=1}^{m} f_j(x_{ij})\right)^2 + \sum_{j=1}^{m} \lambda_j \int f_j''(z)^2 dz$$

where the $\lambda_j > 0$ are tuning parameters, which determine the weight given to the smoothness of each function. The $x_{ij}$ are the knots for $f_j(\cdot)$ and $N$ is the total number of values of the response and factor variables.

We compared the results obtained from this GAM to those of a GLM,[3] which has the general form

$$g(E_Y(Y|X)) = \beta_0 + \beta_1 x_1 + \cdots + \beta_m x_m \equiv X\beta, \qquad (2)$$

where: $Y = [Y_1, \ldots, Y_N]^T$ is the column vector of values of the response variable; $x_j = [x_{j1}, \ldots, x_{jN}]^T$ is the column vector of values for factor $x_j$; $\beta = [\beta_0, \beta_1 \ldots, \beta_m]^T$ is the column vector of unknown parameters; and the columns of $N \times (m+1)$ matrix $X$ correspond to the factor column vectors, except for the first which is the column vector of ones. As we employed the identity link function, (2) becomes a pure linear model, which we refer to as GLM-l.

We also considered GLMs with higher order terms, specifically: <u>GLM-lm</u> having linear and multiplicative factor terms,

$$g(E_Y(Y|X)) = \beta_0 + \sum_{i=1}^{m} \beta_i x_i + \sum_{i=1}^{m-1}\sum_{j=i+1}^{m} \beta_{ij} x_i x_j \ ; \qquad (3)$$

GLM-lq with linear and quadratic factor terms,

$$g(E_Y(Y|X)) = \beta_0 + \sum_{i=1}^{m} \beta_i x_i + \sum_{i=1}^{m} \beta_{ii} x_i^2 \ ; \qquad (4)$$

GLM-lmq with linear, multiplicative, and quadratic factor terms,

---

[3] We utilized the *fitglm* function from MatLab for the implementation of (2). The link function was set to the identity.

$$g\left(E_Y(Y|X)\right) = \beta_0 + \sum_{i=1}^{m} \beta_i x_i + \sum_{i=1}^{m-1} \sum_{j=i+1}^{m} \beta_{ij} x_i x_j + \sum_{i=1}^{m} \beta_{ii} x_i^2 ; \qquad (5)$$

and GLM-p, a polynomial with all factor terms up to the third degree,

$$\begin{aligned} g\left(E_Y(Y|X)\right) = \beta_0 &+ \sum_{i=1}^{m} \beta_i x_i + \sum_{i=1}^{m-1} \sum_{j=i+1}^{m} \beta_{ij} x_i x_j + \sum_{i=1}^{m} \beta_{ii} x_i^2 \\ &+ \sum_{i=1}^{m-2} \sum_{j=i+1}^{m-1} \sum_{k=j+1}^{m} \beta_{ijk} x_i x_j x_k + \sum_{i=1}^{m-1} \sum_{j=i+1}^{m} \left(\bar{\beta}_{ij} x_i^2 x_j + \hat{\beta}_{ij} x_i x_j^2\right) \\ &+ \sum_{i=1}^{m} \beta_{iii} x_i^3 . \end{aligned} \qquad (6)$$

The GLM models (3)-(6) were implemented using the MatLab function *stepwiseglm*. To reduce the number of coefficients in each model, beginning with the constant term, successive terms were added to the model only if the deviance is reduced as a result of the addition. In all cases, the identity link function was utilized.

## 3. Results

### *3.1 Comparison of GAM and GLM-l*

Table 1 presents the significance ($p$-value) for the various factor coefficients as fit by the GAM model (1) and GLM-l (2) for NYC, LA and Louisville. We first contrast the GAM results by city. For the two largest cities in the U.S., all factors considered are very significant ($p \leq 0.005$), with the exception that days on the market is much less significant for NYC. In contrast, for the smaller city of Louisville, Dwelling, Beds and Distance lacked competitive significance. These results are not surprising given that different desirability factors affect large and mid-size urban areas. In contrast, under the linear GLM model, the year of construction is deemed not significant for all cities (significant at only the 5% level for Louisville). For NYC and LA, two additional factors (Dwelling and Distance for NYC, Lot and Days for LA) that were deemed significant under GAM are not significant under the GLM-l model, while for Louisville, an additional four (Dwelling, Latitude, Beds, and Distance) are not significant.

Under the GAM model, the factors considered accounted for 84% to 89% of the log-price variation (adjusted $R^2$ values); under the GLM-l model, adjusted $R^2$ values varied from 67% to 79%. The factor significances, combined with $R^2$ values, recommend the use of GAM pricing models over a basic GLM-l.

The results confirm the significance of using geospatial latitudinal and longitudinal coordinates in considering home prices. It is important to recognize that such a relationship can clearly be nonlinear. Louisville is an excellent example of this. The northwestern part of the city is associated with lower

**Table 1:** Significance (p-value) of the factors in the GLM-l and GAM pricing models

| Factor | p-value | | | | | |
|---|---|---|---|---|---|---|
| | NYC | | LA | | Louisville | |
| | GLM-l | GAM | GLM-l | GAM | GLM-l | GAM |
| Dwelling | 0.564 | * | * | * | 0.106 | 0.0629 |
| Borough | * | * | N/A | N/A | N/A | N/A |
| Latitude | * | * | * | * | 0.976 | 0.00345 |
| Longitude | * | * | * | * | * | * |
| Beds | $3.19 \cdot 10^{-10}$ | * | * | * | 0.584 | 0.320 |
| Baths | * | * | * | * | $5.71 \cdot 10^{-14}$ | 0.000359 |
| Indoors | * | * | * | * | $5.24 \cdot 10^{-6}$ | * |
| Lot | $1.82 \cdot 10^{-12}$ | * | 0.5033 | * | 0.00368 | * |
| Year | 0.532 | * | 0.1707 | * | 0.0311 | * |
| Days | $2.63 \cdot 10^{-5}$ | 0.011 | 0.0898 | $8.14 \cdot 10^{-5}$ | $1.81 \cdot 10^{-5}$ | 0.00188 |
| HOA | * | * | * | * | 0.000277 | $1.18 \cdot 10^{-5}$ |
| Distance | 0.523 | * | $2.09 \cdot 10^{-11}$ | * | 0.599 | 0.921 |
| Adj. $R^2$ | 0.785 | 0.891 | 0.7314 | 0.874 | 0.6703 | 0.837 |

* Indicates $p$-value $< 2 \cdot 10^{-16}$.

home prices, whereas the eastern part of the city is associated with higher home prices. Hence, the GLM-l and GAM models both identify longitude as a significant factor. As the "old city" has a higher latitude than the wealthy area of Spring Mill but a lower latitude than that of Prospect, only the greater flexibility provided by the GAM is able to account for the price variability with latitude.

*3.2 Comparison of GAM with polynomial based GLM models*

Using the data for NYC, the performance of the P-spline GAM model (1) and the linear model GLM-l (2) was compared to the polynomial-based GLM models (3)-(6). Table 2 compares the results based on adjusted $R^2$, mean square error (MSE), the mean absolute relative error (MARE), and the Bayesian information criterion (BIC) of the regression fits. The P-spline GAM model provides superior values in all four goodness-of-fit categories. Of the GLM models, the performance (highest adjusted $R^2$; smallest values of MSE, MARE and BIC) of polynomial model with the most degrees of freedom, GLM-p, is the best.

In NYC, the spatial distribution of condominiums is not as uniform as the other three dwelling types, being more concentrated to Manhattan, the northern neighborhoods of Brooklyn, and the western neighborhoods of Queens (Fig. 2). As a consequence, there is also a noticeable difference in the ln(Price/sq ft) between condominiums and the other dwelling types (Fig. 3). It makes sense therefore to run separate GAM models on the two dwelling classes. We refer to these models as GAM-cond (condominium data) and GAM-non (non-condominium data). The adjusted $R^2$, MSE, MARE, and BIC for these two GAM models are also reported in Table 2. With less diverse data sets, the GAM-non and GAM-cond models would be expected to outperform the GAM model that comprises both data sets. GAM-non outperforms GAM in all but the adjusted $R^2$ measure. GAM-cond outperforms both GAM and GAM-non in all four fit measures. Indeed, the GAM-cond model explains over 92% of the variation in condominium log-price for NYC.

Table 2. Summary fit statistics for GLM and GAM models for NYC

| Model | Adj. $R^2$ | MSE | MARE | BIC |
|---|---|---|---|---|
| GLM-l | 0.774 | 0.1296 | 0.2878 | 6179 |
| GLM-lm | 0.842 | 0.0901 | 0.2579 | 3625 |
| GLM-lq | 0.820 | 0.1030 | 0.2579 | 4479 |
| GLM-lmq | 0.858 | 0.0809 | 0.2212 | 2914 |
| GLM-p | 0.873 | 0.0724 | 0.2074 | 2054 |
| GAM | 0.891 | 0.0625 | 0.1935 | 1181 |
| GAM-non | 0.855 | 0.0569 | 0.1794 | 473 |
| GAM-cond | 0.924 | 0.0495 | 0.1729 | −137 |

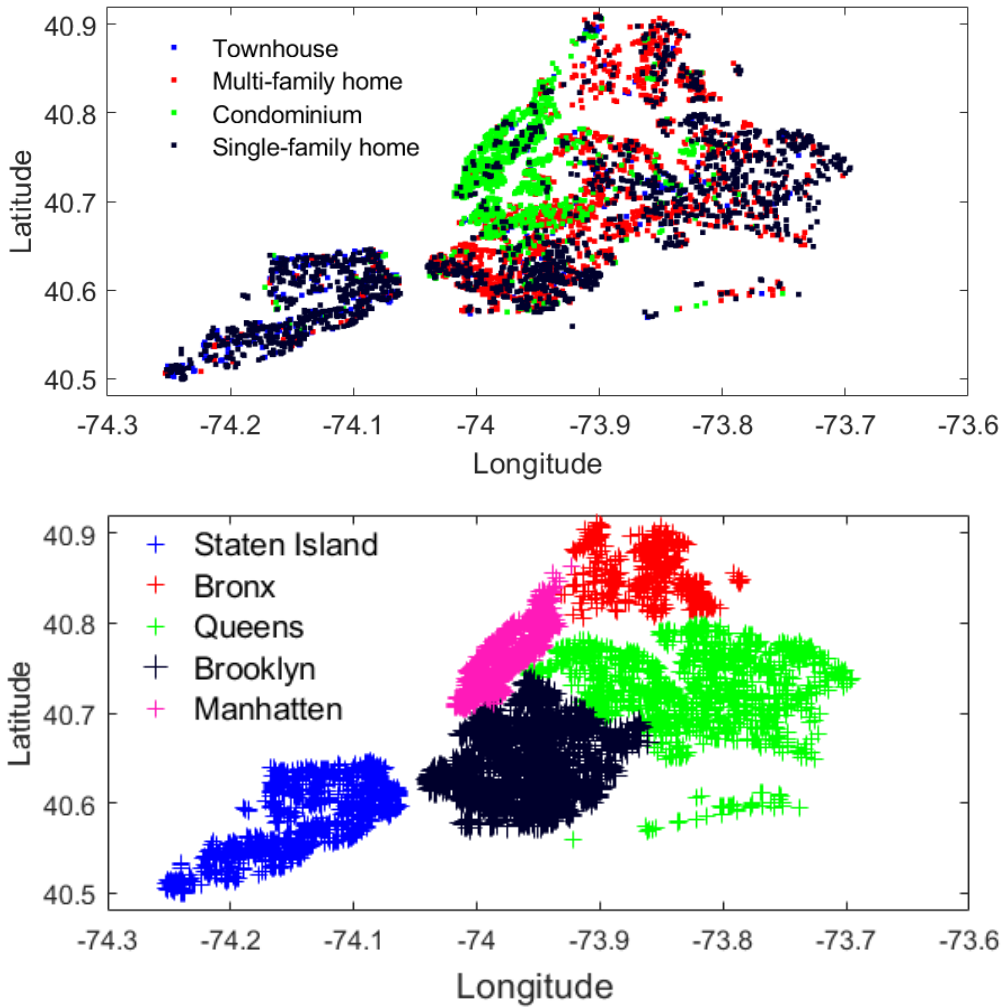

**Figure 2.** The spatial distribution of dwellings in NYC: (top) by type, (bottom) by borough.

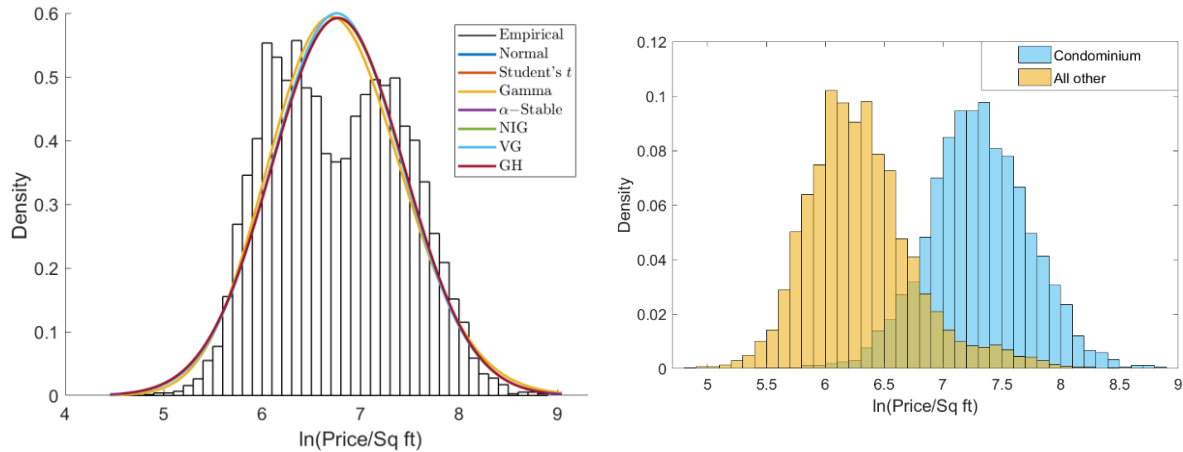

**Figure 3.** The distribution of ln(Price/Area) for (left) all property types in NYC and (right) condominiums plotted separately from the remaining three property types.

*3.3 Inclusion of environmental factors*

Four environmental factors are available in the Redfin data. These are:
- Waterfront - the residence borders or overlooks a body of water;
- Accessible - the residence is accessible by disabled persons;
- Green - green-energy sources (e.g., solar panels) are present in the residence; and
- Air Cond - the residence has an air conditioning unit.

All four environmental factors are Boolean valued. To evaluate the impact of these environmental factors on pricing models, we concentrated on data from cities that have more stringent environmental policies, specifically Portland, OR and Seattle. As the GAM models proved to be the most predictive, for each city we compare the effectiveness of running the GAM model (1) with environmental factors (GAM-env) and without (GAM).[4]

We consider the GAM model first. For Seattle, nine factors were found to be very significant ($p \leq 0.005$), the exceptions being HOA and Distance (although Distance is significant at the 1% level). For Portland, seven factors were very significant, with exceptions being Beds, Baths, Days and Distance (although Beds is significant at the 1% level). For Seattle, when the four environmental factors are added to the GAM model, three of the four (the exception being Accessibility) become very significant, while the significance of Days decreases to the 1% level. In contrast for Portland, of the four environmental factors, only Air Cond is deemed very significant. For both cities, inclusion of the environmental factors increases the adjusted $R^2$ value of the fit by less than 1%.

---

[4] GLM-l models were also run for Seattle and Portland. These fits resulted in adjusted $R^2$ values in the range 72% to 78%.

Table 3. Significance ($p$-value) of GAM pricing model factors with and without the inclusion of environmental factors

| Factor | $p$-value | | | |
|---|---|---|---|---|
| | Portland | | Seattle | |
| | GAM | GAM-env | GAM | GAM-env |
| Dwelling | $7.5 \cdot 10^{-5}$ | 0.002 | * | * |
| Latitude | * | * | * | * |
| Longitude | * | * | * | * |
| Beds | 0.007 | 0.032 | * | * |
| Baths | 0.23 | 0.25 | $1.9 \cdot 10^{-6}$ | $4.0 \cdot 10^{-7}$ |
| Indoors | * | * | * | * |
| Lot | * | * | $2.2 \cdot 10^{-7}$ | $8.9 \cdot 10^{-6}$ |
| Year | * | * | * | $7.5 \cdot 10^{-6}$ |
| Days | 0.88 | 0.46 | 0.004 | 0.010 |
| HOA | 0.005 | 0.11 | 0.084 | 0.069 |
| Distance | 0.042 | 0.12 | 0.009 | 0.048 |
| Waterfront | NI | 0.36 | NI | $1.1 \cdot 10^{-10}$ |
| Accessible | NI | 0.019 | NI | 0.42 |
| Green | NI | 0.013 | NI | $1.6 \cdot 10^{-4}$ |
| Air Cond | NI | $1.2 \cdot 10^{-6}$ | NI | 0.002 |
| Adj. $R^2$ | 0.875 | 0.884 | 0.871 | 0.879 |

\* Indicates $p$-value $< 2 \cdot 10^{-16}$.    NI = not included in model

## 4. Discussion

Our results demonstrate that P-spline GAM hedonic models have very strong predictive capability (adjusted $R^2$ values in the range 84% to 92%) for the expected value of the ln(sale price) of residence units in major U.S. cities. This contrasts to linear models (GLM-l) with adjusted $R^2$ in the range 65% to 78%. Use of polynomial based GLMs improved adjusted $R^2$ values to the range 82% to 88%, but did not outperform the P-spline GAM. The high $R^2$ values obtained for GAM imply that other microeconomic or macroeconomic factors not included in our study account for less than 15% of the variance in housing price.

The results confirm the importance of including latitude and longitude as factors. These are critical proxies for the "location, location, location" real estate axiom reflecting the existence of desirable school district, neighborhoods, etc. The results related to the significant of the distance to the nearest sex offender as a hedonic factor are mixed, with indications that this is very significant in NYC and LA, but less so in the other three cities considered. A deeper consideration of city/state policies regulating sex offender residence location is required to understand these results. In New York State, the Sex Offender Registration Act does not restrict where a registered sec offender may live. In California, blanket restrictions imposed under Jessica's Law were invalidated by the state supreme court and residency restrictions are evaluated on a case-by-case basis. In Washington State, "sex offenders are explicitly prevented from living in a residence that is proximate to a school, child care center, playground or other facility where children of a similar age or circumstances as a previous victim is present and would be put at substantial risk of harm".[5] Laws similar to Washington State hold in Oregon and Kentucky.

Based upon the two cities studied, we suggest that the significance of environmental factors is still very city dependent. In the U. S., real estate must comply with regulations at the municipal, city, county, city, and federal levels. Such regulations related to environmental factors (hurricane resistance, flood plain location) are increasing under the pressures of climate change. We suggest a study related to residence risk from increasingly occurring natural disasters (hurricanes, floods, wildfires) is called for. A Natural Disasters Index (e.g., Mahanama et al., 2021) which quantifies such financial risk could be included as a factor.

**Appendix. Data**

In accessing data from Redfin, for each city chosen, the specific values used for "All filters" are listed in Table A1.

**Table A1.** Filter settings used in accessing Redfin data

| Filter | Value | Filter | Value |
|---|---|---|---|
| **Price** | Min: $50K, Max: $10M | **Beds** | 1+ |
| **Home type** | House, Townhouse, Condo, Multi-family | **Baths** | 1+ |
| **Status** | Coming soon, Active | **Property details** | |
| Under contract/pending | No | Square feet | Min: 250, Max: NS |
| | | Lot size | Min: 250, Max: NS |
| **Time on Redfin** | No max | Stories | Min: NS, Max: NS |
| Exclude 55+ communities | No | Year built | Min: NS, Max: NS |
| **Home features** | | | |
| Garage Spots | Any | Pool type | Any |
| Include outdoor parking | No | Basement | NS [1] |
| Air conditioning | ESG [2] | Waterfront | ESG [2] |
| Washer/dryer hookup | NS | Has view | NS |
| Pets allowed | NS | Fireplace | NS |
| Master B/R on main floor | NS | Fixer-upper | NS |
| RV parking | NS | Guest house | NS |
| Green home | ESG [2] | Elevator | NS |

---

[5] Review of Policies Relating to the Release and Housing of Sex Offenders in the Community, December 2014. Sex Offender Policy Board, Office of Financial Management, State of Washington. https://sgc.wa.gov/sites/default/files/public/sopb/documents/sex_offender_housing_201412.pdf

| | | | | |
|---|---|---|---|---|
| Accessible home | ESG [2] | | | |
| Keyword search | NS | | | |
| | | **Cost/finance** | | |
| HOA fees | No max | | Property taxes | No max |
| Price/Sq ft | Min: NS | Max: NS | Accepted financing | NS |
| Exclude land leases | No | | Price reduced | No |
| | | **Listing type** | | |
| By agent | Yes | | Foreclosures | Yes |
| By owner (FSBO) | Yes | | Exclude short sales | No |
| New construction | Yes | | Redfin listing only | No |
| **Schools** | NS | | | |
| **Open Houses & Tour** | NS | | **Walk Score** | NS |

[1] NS = Not specified   [2] Specified only for environmental inclusion